# Controlled layer-by-layer oxidation of MoTe$_2$ via O$_3$ exposure

Xiaoming Zheng[1,3,#], Yuehua Wei[2,#], Chuyun Deng[1], Han Huang[3], Yayun Yu[1], Guang Wang[1], Gang Peng[1], Zhihong Zhu[2], Yi Zhang[1], Tian Jiang[2], Shiqiao Qin[2], Renyan Zhang[2,*], Xueao Zhang[1,3,*]

[1]College of Arts and Science, National University of Defense Technology, Changsha, Hunan, 410073, China
[2]Academy of Advanced Studies in Interdisciplinary Research, National University of Defense Technology, Changsha, Hunan, 410073, China
[3]School of Physics and Electronics, Central South University, Changsha, Hunan, 410004, China

**Abstract:**
Growing uniform oxides with various thickness on TMDs is one of the biggest challenges to integrate TMDs into complementary metal oxide semiconductor (CMOS) logic circuits. Here, we report a layer-by-layer oxidation of atomically thin MoTe$_2$ flakes via ozone (O$_3$) exposure. The thickness of MoO$_x$ oxide film could be tuning with atomic-level accuracy simply by varying O$_3$ exposure time. Additionally, MoO$_x$-covered MoTe$_2$ shows a hole-dominated transport behavior. Our findings point to a simple and effective strategy for growing homogenous surface oxide film on MoTe$_2$, which is promising for several purposes in metal-oxide-semiconductor transistor, ranging from surface passivation to dielectric layers.

Thin oxide layer is critical for the development of modern semiconductor manufacturing technology, which has been widely adopted for insulating in active devices, for physical masking in patterning process, for passivation to protect function materials from contamination[1]. $SiO_2$ or atomic-layer-deposited (ALD) $Al_2O_3$ has been widely adopted as dielectric or insulating layers in transitional metal dichalcogenides (TMDs)-based complementary metal oxide semiconductor (CMOS) devices[2, 3]. However, the mismatch of lattice parameters between oxide layers and TMDs flakes induces conspicuous mechanical stress in these devices[4]. Moreover, physical bombardment during deposition process would result in non-ignorable crystalline damage in TMDs flakes[5]. In consideration of the atomic thickness of TMDs flakes, these issues would seriously degrade the electronic performance and further limit the realist application of TMDs. Thus, growing high quality oxide film with well-controlled interface to underneath TMDs lattice is high desirable to improve the performance of TMDs based device. Recently, it was found that transition metal oxides (TMOs) have good compatibilities with TMDs[6]. Thus, TMOs are expected to be unequalled oxide layers on TMDs, similar with $SiO_2$ on Si in silicon technologies.

However, due to the absence of dangling bonds[7], pristine atomically smooth surface of TMDs is usually resistant to oxidation[8]. For example, in $MoS_2$, oxygen exposure leads to triangular etch pits on the surface even at high temperature 340 ℃, i.e. no oxide film is formed [9, 10]. While for transition metal selenide materials, such as $MoSe_2$[11] and $WSe_2$[11-13], $O_3$ exposure or remote oxygen plasma results in a homogenous surface oxide film. However, the oxidation is self-limiting to the top first layer at room temperature and top three layers even at 200 ℃. This is clearly not sufficient to be used in CMOS device fabrication. While, for transition metal tellurium materials, the oxidation process remains an open question. In consideration of the relative stability of sulfide, selenide and tellurium materials, transition metal tellurium materials are considered as a possible platform to grow high-quality oxides with well-controlled interface and various thickness, as silicon in semiconductor technologies. This will be of great practical significance to integrate TMDs with CMOS logic circuits for future very-large scale integration (VLSI) fabrication.

In this report, we demonstrate the controlled layer-by-layer oxidation of $MoTe_2$ via $O_3$ exposure (Figure 1a), where a homogeneous atomically flat oxide film of non-stoichiometric molybdenum oxides ($MoO_x$ with x<3) forms. The oxides are nucleated from edges and surface defects of the-top first layer and grow laterally. Further $O_3$ exposure, oxidation evolves in the layer-by-layer regime. By varying $O_3$ exposure time, we were able to tune the thickness of $MoO_x$ oxide film, different from the self-limiting oxidation in $WSe_2$[12, 13]. Surprisingly, the crystal quality of underlying unoxidized $MoTe_2$ is found to be barely affected by $O_3$ treatment, which are found to be p-type doped due to the high working function of $MoO_x$. Our findings propose an unconventional but simple approach to grow homogenous oxide on $MoTe_2$, which also pave the way for further research of the mechanism of oxidation in other TMDs.

Figure 1b, c shows the optical images of thin $MoTe_2$ flakes before and after $O_3$ treatment for 2 mins, respectively. It is obvious that $O_3$ treatment cause the change in the optical color contrast. More specifically, the oxidized monolayer become almost invisible, while oxidized

bilayer turns to be similar with pristine monolayer. The change of optical color contrast implies that only the top single layer is oxidized by O₃ exposure for 2 mins. The SEM images (Figure S2) show that the surface morphology remains unchanged after O₃ treatment, which indicates that oxidation results in a homogeneous oxide film. While, the AFM height profiles (Figure 1b, c inset) show that the thickness of monolayer MoTe₂ increase from 1 nm to 1.7 nm after O₃ exposure for 2 mins. Additionally, the surface root-mean-square (rms) roughness of the oxide film is about 0.3 nm, which is nearly atomically smooth[13].

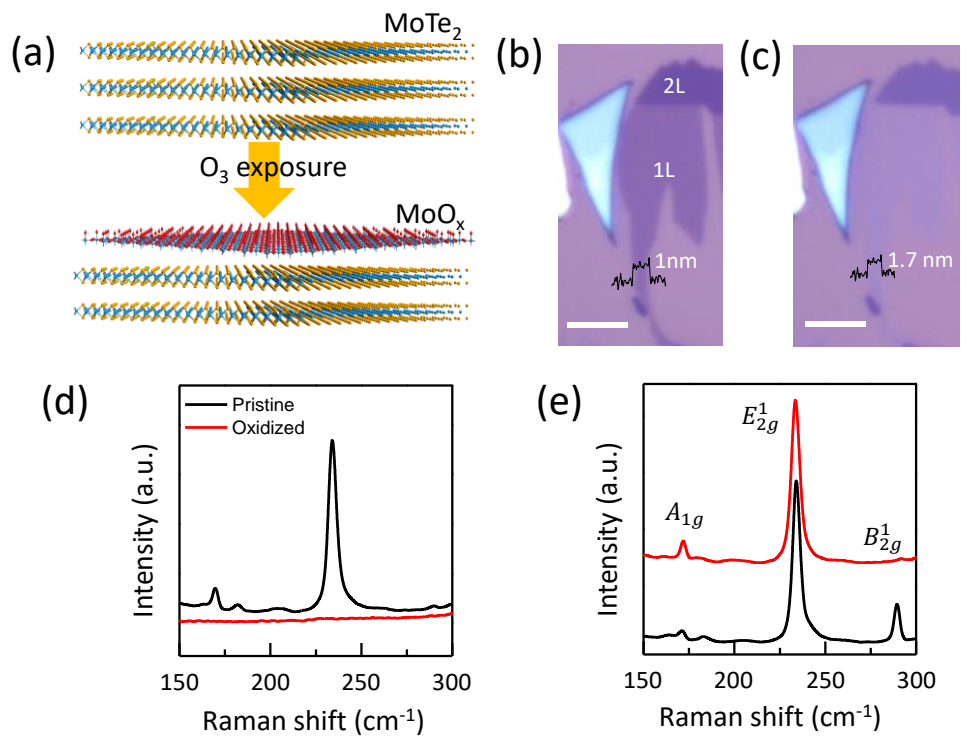

**Figure 1| Oxidation of MoTe₂ by O₃ exposure.** (a) Schematic of atomically thin MoTe₂ flake before and after O₃ exposure for 2 mins. (b, c) Optical image of atomically thin MoTe₂ flakes before and after O₃ exposure for 2 mins, respectively. The scale bar is 5 μm. The inset shows the corresponding height profiles. (d, e) Raman spectra of monolayer (c) and bilayer (d) MoTe₂ before (black curves) and after (red curves) O₃ exposure for 2 mins with an exciting laser wavelength of 532 nm.

Figure 1d, e and S3a, b show the Raman scattering of mono- and bilayer MoTe₂ before and after O₃ exposure for 2 mins. After O₃ treatment for 2 mins, all the Raman scattering modes are vanishing in monolayer MoTe₂, which indicates no residual crystalline MoTe₂ structures after oxidation. However, in case of bilayer MoTe₂, only $B_{2g}^1$ mode disappears. This suggests that the bilayer MoTe₂ becomes monolayer[14, 16], with the top-first layer being oxidized. These results are consistent with the change of optical color contrast as discussed above. It should be noted that there are no Raman modes for oxide film, which is probably because the oxide film is under-stoichiometry and amorphous rather than crystalline. Furthermore, the Raman mapping (Figure S5) of monolayer MoTe₂ upon intermittent O₃ exposure from 0 to 120 s, shows that MoOₓ (with x≤3) oxides are nucleated from the edges and surface defective sites and grow laterally, similar to that in WSe₂ [13, 17] and MoS₂ [10, 18]. After O₃ exposure for 2 mins, the oxide regions coalesce and oxidation terminates leaving a homogenous oxide film. Surprisingly, we found that the full width at half maxima (FWHM) of the Raman $E_{2g}^1$ mode of bilayer MoTe$_2$ only slight increase from 5.2 cm$^{-1}$ to 5.4 cm$^{-1}$ after O$_3$ exposure for 2 mins, which indicates that the oxidation does not compromise the crystalline quality of the underneath unoxidized MoTe$_2$ layer. Above all, O$_3$ exposure is an unequal way to growth oxides with few defects both in the bulk and at the interface between MoO$_x$ and MoTe$_2$.

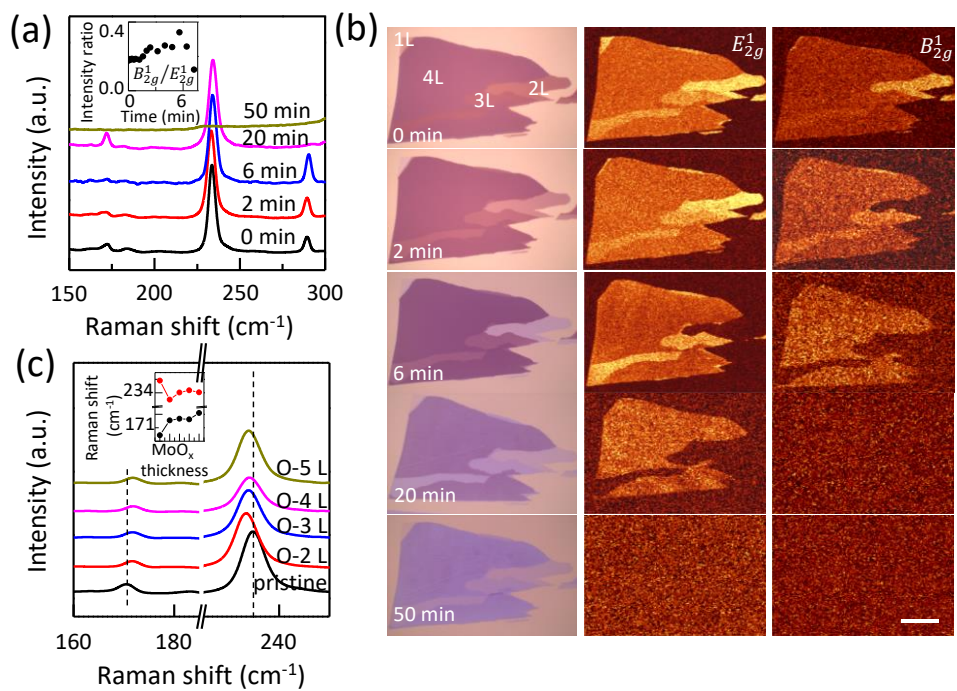

**Figure 2| Layer-by-layer oxidation of few layer MoTe$_2$.** (a) Raman spectra (laser wavelength 532nm) of tetralayer MoTe$_2$ upon intermittent O$_3$ exposure from 0 to 50 mins. The inset shows the relative intensity of $B_{2g}^1/E_{2g}^1$. (b) The corresponding Raman mapping of $E_{2g}^1$ and $B_{2g}^1$ modes. The scale bar is 5 μm. (c) Raman spectra for monolayer and oxidized bilayer, trilayer, tetralayer and five-layer MoTe$_2$ samples. The oxidized n-layer MoTe$_2$ indicates the single layer MoTe$_2$ produced by oxidation of n-layer counterpart. The inset shows the peak position shift of the $E_{2g}^1$ and $A_{1g}$ modes.

Figure 2a, b presents the Raman spectra and corresponding $E_{2g}^1$ and $B_{2g}^1$ modes mapping of tetralayer of MoTe$_2$ flakes for intermittent O$_3$ exposure from 0 to 50 mins. The intensity ratio of $B_{2g}^1/E_{2g}^1$ (Figure 2a inset) is found to increase to the maximum with O$_3$ exposure time increase to 6 mins. After that, the relative intensity gradually decreases to 0 until 20 mins. This phenomenon indicates the reduction of MoTe$_2$ layer thickness with O$_3$ exposure[14, 15], which decrease to bilayer at 6 mins and monolayer at 20 mins. The Raman mapping (Figure 2b) shows that after oxidation for 2 mins, monolayer MoTe$_2$ shows neither $E_{2g}^1$ or $B_{2g}^1$ mode, while bilayer MoTe$_2$ exhibits only the $E_{2g}^1$ mode. This implies that the top single layer is oxidized. Additional 4 mins exposure leads to the vanish of the $B_{2g}^1$ mode in trilayer not the tetralayer MoTe$_2$ flakes, i.e. the top two layers are oxidized. After oxidation for 20 mins, tetralayer MoTe$_2$ shows only a prominent $E_{2g}^1$ peaks, while the $B_{2g}^1$ mode is absent, indicating the top three layers are oxidized with a single layer remaining underneath. Further

30 mins exposure to O$_3$ leads to the oxidation of the fourth layer. The thickness of oxide layer could be tuning simply by changing O$_3$ exposure time with atomic-level accuracy, which is more advantageous for realistic applications than the self-limiting oxidation process in WSe$_2$[13]. For example, the oxide film could sever as a passivation layer to protect MoTe$_2$ from contamination (Figure S7) and as a seed layer to grow high quality A$_2$O$_3$ film (Figure S8). Additionally, we investigate O$_3$ oxidation of MoTe$_2$ flakes at higher temperatures (Figure S9), which shows that the oxidation processes at 150 °C and 200 °C are still layer-by-layer regime. However, the oxidation speed is increase dramatically at high temperature.

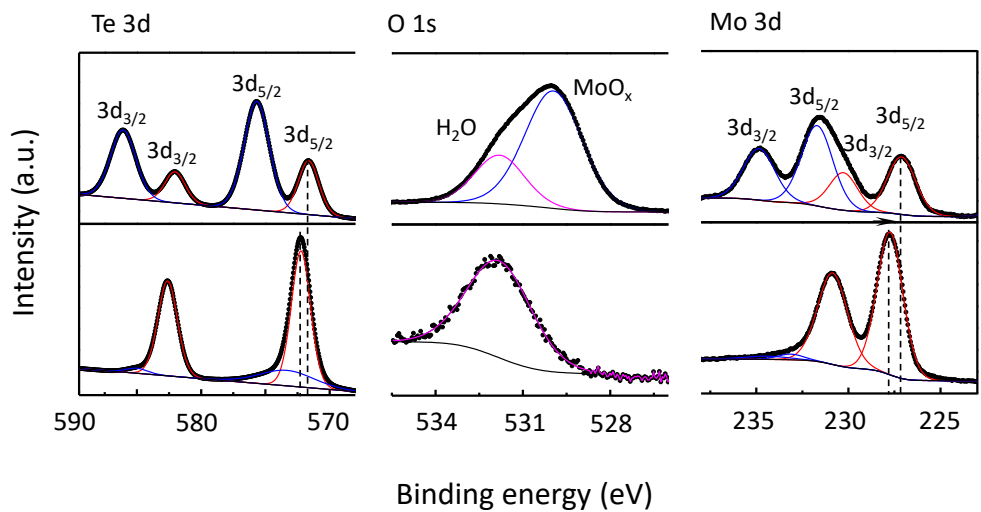

**Figure 3| XPS spectra of bulk MoTe$_2$ before and after O$_3$ exposure.** The shown energy region corresponds the binding energy of Te 3d, O 1s and Mo 3d core level. The lower and upper panel are the XPS spectra before and after O$_3$ exposure for 60 mins, respectively. Black dots are experimental data. Red and blue curves are the corresponding Lorentzian fits to Te 3d, O 1s and Mo 3d peaks for MoTe$_2$ and MoO$_x$, respectively. The magenta curve is the Lorentzian fits to O 1s peak for absorbed H$_2$O.

Figure 3 shows the XPS spectra of Te 3d, O 1s and Mo 3d core level of bulk MoTe$_2$ sample before and after O$_3$ exposure for 60 mins. All spectra were calibrated using C 1s peak at 284.6 eV as a reference. Pristine MoTe$_2$ shows Mo 3d$_{5/2}$ and Mo 3d$_{3/2}$ doublet features at 227.8 and 230.9 eV, respectively. After O$_3$ exposure for 60 mins, XPS spectrum shows the emergence of two new peaks at higher binding energy of 231.7 and 234.8 eV, which corresponds to 3d$_{5/2}$ and 3d$_{3/2}$ peaks of nonstoichiometric molybdenum oxide (MoO$_x$ with x<3)[19, 20]. The O 1s core level spectra also suggest that the oxidation product is MoO$_x$ [19]. Moreover, the doublet originating from Mo 3d core level of MoTe$_2$ shifts toward lower binding energy by 0.7 eV, compared with the spectrum of pristine MoTe$_2$. Similar behaviors are also found in Te 3d core level spectra. The downshift effect indicates the presence of a p-type doping effect after O$_3$ treatment[21, 22]. Raman peaks of single layer MoTe$_2$ produced by oxidation of n-layer counterpart (named as oxidized n-layer MoTe$_2$) also show peak position shifts, when compared with that of pristine monolayer MoTe$_2$ (Figure 2c). $E_{2g}^1$ mode of single layer MoTe$_2$ with oxide is soften by 1.2 cm$^{-1}$ after oxidation. While, $A_{1g}$ mode shows an upshift from 170.4 cm$^{-1}$ to 171.7 cm$^{-1}$. This is another indication of p-type doping after O$_3$ exposure.

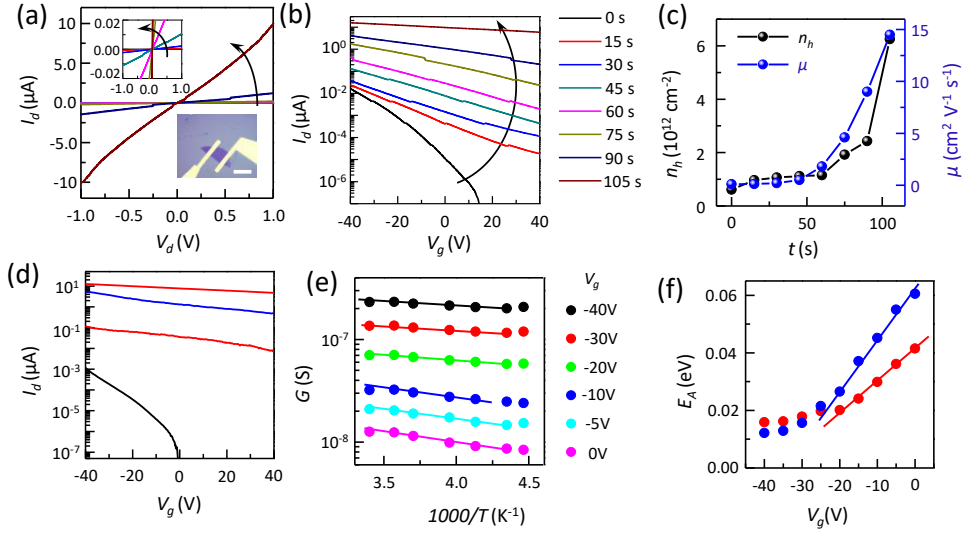

**Figure 4| Transport properties of bilayer MoTe$_2$ upon intermittent O$_3$ exposure from 0 to 105 s.** (a, b) $I_d$ - $V_d$ and $I_d$ – $V_g$ characteristics of bilayer MoTe$_2$ transistor upon intermittent O$_3$ exposure from 0 to 105 s, respectively. (c) The two-probe hole FET mobility and carrier density as a function of O$_3$ exposure time. (d) Transfer curves of monolayer (black curve), oxidized bilayer (red curves for two different samples) and oxidized trilayer MoTe$_2$ (blue curve) devices at $V_d$ =1 V. (e) Arrhenius plots of the two-probe conductivity (G) of oxidized bilayer MoTe$_2$ for various $V_g$. (f) Total activation energy $E_A$ as a function of $V_g$ of oxidized bilayer (red dots) and oxidized trilayer (blue dots) MoTe$_2$ devices. The red and blue lines are the linear fits.

To gain more insight into the doping properties of O$_3$ treated MoTe$_2$, we carry out transport measurements on a back-gate bilayer MoTe$_2$ FTE (Figure 4a inset) before and after O$_3$ intermittent exposure from 0 to 105 s. Figure 4a presents the $I_d$ versus $V_d$ curves of bilayer MoTe$_2$ FETs under various O$_3$ exposure time, where $I_d$ keeps increase with the increase of O$_3$ exposure time. The corresponding logarithmic plots of transfer characteristics ($I_d$ - $V_g$) at $V_d$ = 1 V are shown in Figure 4b. $I_d$ of pristine bilayer MoTe$_2$ under 1 V bias increased only along the negative gate voltage region, corresponding to the hole dominated transport behavior. This agrees with previous observation in MoTe$_2$ [26]. The transfer curve kept positively shifting with longer O$_3$ exposure time, which indicates the increase of p-type doping level. After O$_3$ exposure for 105 s, MoTe$_2$ device reveals degenerate-like transfer characteristics. By varying O$_3$ exposure time, the hole mobility $\mu_h$ of oxidized MoTe$_2$ transistors increase from 0.104 to 14.5 cm$^2$/(V s) at $V_g$ = -40 V, which is comparable to the highest mobilities ever obtained in MoTe$_2$ transistors at room temperature [22]. In the meantime, the hole doping concentration could be modulated over a wide range, from slight p-type doping (6.1 x 10$^{11}$ cm$^{-2}$) to degenerate-like doping (6.3 x 10$^{12}$ cm$^{-2}$) at $V_g$ = -40 V. The relationship of mobility and hole density with respect to O$_3$ treatment time are summarized in Figure 4c. Similar phenomena are also found in trilayer MoTe$_2$ devices (Figure S12).

The doping level of oxidized n-layer MoTe$_2$ varies from sample to sample with no clear dependence on the thickness of oxide layers (Figure 2c and 4d). This is very different from that of thermally evaporated MoO$_x$ surface dopant[26, 27]. We suggest that oxidation forms a better interface with MoTe$_2$, compared to thermally evaporated MoO$_x$. The physical bombardment during thermal evaporation may cause crystalline damages on the surface of MoTe$_2$, resulting in a rough interface, which contribute to additional contact resistance. In contrast, oxidation leads to a lower degree of defects at the interface, as already been demonstrated by Raman measurement. Additionally, e-beam evaporated TMO films are not continuous below a certain thickness limit [28, 29]. This means the surface doping would not saturate below this critical thickness. While for oxidation, the oxide film is flat and homogeneous even for atomic thin samples. Thus, oxidation product a well-controlled interface between MoO$_x$ and MoTe$_2$, and the p-type doping is saturate even for oxide bilayer MoTe$_2$ sample. The sample dependence doping level may cause by the quality of samples and contacts. To better understanding the contact properties of oxidized MoTe$_2$ transistors, we analysis the activation energy of the oxidized bilayer and trilayer MoTe$_2$ devices. Figure 4e shows the Arrhenius plots of the two-probe conductivity (G ~ $-\frac{E_A}{k_B T})^{12}$ of oxidized bilayer MoTe$_2$ device. The slop of Arrhenius plot is used to extract the total activation energy ($E_A$) at certain $V_g$, which is summarized in Figure 4f. At $V_g$ > -20 V, the $E_A$ - $V_g$ curves are linear. Thus, we can extract the Schottky barrier height of oxidized bilayer and trilayer MoTe$_2$ devices at $V_g$ = -20 V, which are 19 and 23 meV, respectively. This low barrier height is consistent with the increase of mobility of oxidized MoTe$_2$ devices.

In summary, we have demonstrated that O$_3$ treatment provides a layer-by-layer oxidation of MoTe$_2$. O$_3$ exposure leads to the formation of a uniform MoO$_x$ (with x<3) film, which is atomically flat with the rms roughness comparable to that of atomically thin MoTe$_2$ on SiO$_2$. Further O$_3$ exposure results in a layer-by-layer oxidization of MoTe$_2$. The exposure time is adopted to control the thickness of surface oxide film. Raman, XPS and transport measurement shows that surface oxides on atomically thin MoTe$_2$ can be used as effective p-type dopants. Additionally, the surface MoO$_x$ oxide is demonstrated to be a good passivation layer to prevent MoTe$_2$ from contamination and seed layer for high dielectrics. Our findings are the first step toward controllable growth of oxides on layered transition metal dichalcogenides to integrate TMDs in CMOS logical circus.

## ASSOCIATED CONTENT
**Supporting Information:** Experimental section and additional information including the O$_3$ UV absorption spectrum, SEM images, Raman spectra, optical images and transport measurements. And the detail results of the MoTe$_2$ oxidation experiments at higher temperature.

## AUTHOR INFORMATION
**Corresponding Author**
*Renyan Zhang: ryancms@sina.cn
*Xueao Zhang: xazhang@nudt.edu.cn
**ORCID**
Renyan Zhang: 0000-0001-6949-8888

**Author Contributions**

# X.Z and Y.W contributed equally to this work. H.H, R.Z and X.Z conceived the project. X.Z and Y.W prepared and characterized all samples. Y.Y and G.W carried out XPS measurements. R.Z and X.Z supervised the work. X.Z, Y.W, Y.D, Y.Z, H.Z, T.J, G.P, S.Q, H.H, R.Z and X.Z analyzed that data and wrote the manuscript. All authors have given approval to the final version of the manuscript.

**Notes**

The authors declare no competing financial interest.

**ACKNOWLEDGMENT**

This work was supported by the National Natural Science Foundation (NSF) of China (Grant No. 61675234, 11304398, 11574395, 11404399), the Research Project of National University of Defense Technology (No.JC15-02-01), the Advanced Research Foundation of the National University of Defense Technology (Grant No. zk16-03-40), the Driven project of Central South University (No. 2017 CX018), the NSF of Hunan province (Grants No.2016JJ1021), and the Fundamental Research Funds for the Central Universities of Central South University (Grant No. 2017zzts066).

**References**

(1) Sze, S. M., Semiconductor Devices :Physics and Technology, John Wiley & Sons, 2008.

(2) Lin, F.; Xu, Y.; Wang, S. T.; Li, S. L.; Yamamoto, M.; Aparecido-Ferreira, A.; Li, W.; Sun, H.; Nakaharai, S.; Jian, B. Ambipolar $MoTe_2$ Transistors and Their Applications in Logic Circuits. *Adv. Mater.,* 2014, 26, 3263-3269.

(3) Zhang, H.; Chiappe, D.; Meersschaut, J.; Conard, T.; Franquet, A.; Nuytten, T.; Mannarino, M.; Radu, I.; Vandervorst, W.; Delabie, A. Nucleation and Growth Mechanisms of $Al_2O_3$ Atomic Layer Deposition on Synthetic Polycrystalline $MoS_2$. *J. Chem. Phys.,* 2017, 146, 263-275.

(4) Yun, S.; Han, W.; Hong, C.; Kim, G.; Lee, D. Thickness and Strain Effects on Electronic Structures of Transition Metal Dichalcogenides: $2H-MX_2$ Semiconductors (M = Mo, W; X = S, Se, Te). *Phys. Rev., B Condens. Matter* 2012, 85(3), 033305.

(5) Liu, Y.; Guo, J.; Zhu, E.; Liao, L.; Lee, J.; Ding, M.; Shakir, I.; Gambin, V.; Huang, Y.; Duan, X., Approaching the Schottky–Mott Limit in Van Der Waals Metal–Semiconductor Junctions. *Nature* 2018, 557, 696–700.

(6) Mcdonnell, S.; Azcatl, A.; Addou, R.; Gong, C.; Battaglia, C.; Chuang, S.; Cho, K.; Javey, A.; Wallace, M., Hole Contacts on Transition Metal Dichalcogenides: Interface Chemistry and Band Alignments. *ACS Nano* 2014, 8, 6265-6272.

(7) Lin, C.; Ghosh, K.; Addou, R.; Lu, N.; Eichfeld, M.; Zhu, H.; Li, Y.; Peng, X.; Kim, J.; Li, J., Atomically Thin Resonant Tunnel Diodes Built From Synthetic Van Der Waals Heterostructures. *Nat. Commun.* 2015, 6, 7311.

(8) Wang, H.; Kalantarzadeh, K.; Kis, A.; Coleman, N.; Strano, S., Electronics and Optoelectronics of Two-Dimensional Transition Metal Dichalcogenides. *Nat. Nanotechnol.* 2012, 7, 699-712.

(9) Dhall, R.; Neupane, R.; Wickramaratne, D.; Mecklenburg, M.; Li, Z.; Moore, C.; Lake, K.; Cronin, S., Direct Bandgap Transition in Many-Layer $MoS_2$ by Plasma-Induced Layer Decoupling. *Adv. Mater.* 2015, 27, 1573-1578.

(10) Yamamoto, M.; Einstein, L.; Fuhrer, S.; Cullen, G., Anisotropic Etching of Atomically Thin $MoS_2$.

*J. Phys. Chem. C* 2013, 117, 25643–25649.

(11) Zhen, L.; Sisi, Y.; Rohan, D.; Ewa, K.; Haotian, S.; Ioannis, C.; Stephen, C., Layer Control of WSe$_2$ via Selective Surface Layer Oxidation. *ACS Nano* 2016, 10, 6836-6842.

(12) Yamamoto, M.; Nakaharai, S.; Ueno, K.; Tsukagoshi, K. Self-Limiting Oxides on WSe$_2$ as Controlled Surface Acceptors and Low-Resistance Hole Contacts. *Nano Lett.* 2016, 16(4), 2720-2727.

(13) Yamamoto, M.; Dutta, S.; Aikawa, S.; Nakaharai, S.; Wakabayashi, K.; Fuhrer, M. S.; Ueno, K.; Tsukagoshi, K. Self-Limiting Layer-by-Layer Oxidation of Atomically Thin WSe$_2$. *Nano Lett.* 2015, 15(3), 2067-2073..

(14) Lezama, I.; Arora, A.; Ubaldini, A.; Barreteau, C.; Giannini, E.; Potemski, M.; Morpurgo, A. F. Indirect-to-Direct Band-Gap Crossover in Few-Layer MoTe$_2$. *Nano Lett.* 2015, 15(4): 2336-2342.

(15) Yamamoto, M.; Wang, T.; Ni, M.; Lin, F.; Li, L.; Aikawa, S.; Jian, B.; Ueno, K.; Wakabayashi, K.; Tsukagoshi, K., Strong Enhancement of Raman Scattering from a Bulk-Inactive Vibrational Mode in Few-Layer MoTe$_2$. *ACS Nano* 2014, 8, 3895-3903.

(16) Ruppert, C.; Aslan, B.; Heinz, F., Optical Properties and Band Gap of Single- and Few-Layer MoTe$_2$ Crystals. *Nano Lett.* 2014, 14(11), 6231-6236.

(17) Liu, Y.; Tan, C.; Chou, H.; Nayak, A.; Wu, D.; Ghosh, R.; Chang, Y.; Hao, Y.; Wang, X.; Kim, S., Thermal Oxidation of WSe$_2$ Nanosheets Adhered on SiO$_2$/Si Substrates. *Nano Lett.* 2015, 15, 4979-4984.

(18) Jaegermann, W.; Schmeisser, D., Reactivity of Layer Type Transition Metal Chalcogenides Towards Oxidation. *Surf. Sci.* 1986, 165, 143-160.

(19) Moulder, F.; Chastain, J.; King, C., Jr, Handbook of X-Ray Photoelectron Spectroscopy : a Reference Book of Standard Spectra for Identification and Interpretation of XPS Data. *Chem. Phys. Lett.* 1992, 220, 7-10.

(20) Siokou, A.; Leftheriotis, G.; Papaefthimiou, S.; Yianoulis, P., Effect of the Tungsten and Molybdenum Oxidation States on the Thermal Coloration of Amorphous WO$_3$ and MoO$_3$ Films. *Surf. Sci.* 2001, 482: 294-299.

(21) Zhang, R.; Drysdale, D.; Koutsos, V.; Cheung, R., Controlled Layer Thinning and p-Type Doping of WSe$_2$ by Vapor XeF$_2$. *Adv. Funct. Mater.* 2017, 27(41), 1702455..

(22) Qu, D.; Liu, X.; Huang, M.; Lee, C.; Ahmed, F.; Kim, H.; Ruoff, S.; Hone, J.; Yoo, J., Carrier-Type Modulation and Mobility Improvement of Thin MoTe$_2$. *Adv. Mater.* 2017, 29(39): 1606433.

(23) Meyer, J.; Hamwi, S.; Kröger, M.; Kowalsky, W.; Riedl, T.; Kahn, A., Transition Metal Oxides for Organic Electronics: Energetics, Device Physics and Applications. *Adv. Mater.* 2012, 24(40), 5408-5427.

(24) Greiner, T.; Lu, H., Thin-Film Metal Oxides in Organic Semiconductor Devices: Their Electronic Structures, Work Functions and Interfaces. *Npg Asia Mater.* 2013, 5(7), 547-556.

(25) Michaelson, H. B., The Work Function of the Elements and Its Periodicity. *J. Appl. Phys.* 1977, 48, 4729-4733.

(26) Luo, W.; Zhu, M.; Peng, G.; Zheng, X.; Miao, F.; Bai, S.; Zhang, A.; Qin, S., Carrier Modulation of Ambipolar Few-Layer MoTe$_2$ Transistors by MgO Surface Charge Transfer Doping. *Adv. Funct. Mater.* 2018, 28(15), 1704539.

(27) Xiang, D.; Han, C.; Wu, J.; Zhong, S.; Liu, Y.; Lin, J.; Zhang, A.; Ping, W.; B, Ö.; Neto, H., Surface Transfer Doping Induced Effective Modulation on Ambipolar Characteristics of Few-Layer Black Phosphorus. *Nat. Commun.* 2015, 6, 6485.

(28) Miyata, N.; Suzuki, T.; Ohyama, R., Physical Properties of Evaporated Molybdenum Oxide Films. *Thin Solid Films* 1996, 281, 218-222.

(29) Yang, Q.; Wei, R.; Zhu, H.; Yang, Y., Strong Influence of Substrate Temperature on the Growth of Nanocrystalline $MoO_3$ Thin Films. *Phys. Lett. A.* 2009, 373, 3965-3968.

(30) Chen, B.; Sahin, H.; Suslu, A.; Ding, L.; Bertoni, I.; Peeters, M.; Tongay, S., Environmental Changes in $MoTe_2$ Excitonic Dynamics by Defects-Activated Molecular Interaction. *ACS Nano* 2015, 9, 5326-5332.

**Figure for the Table of Contents (TOC)**

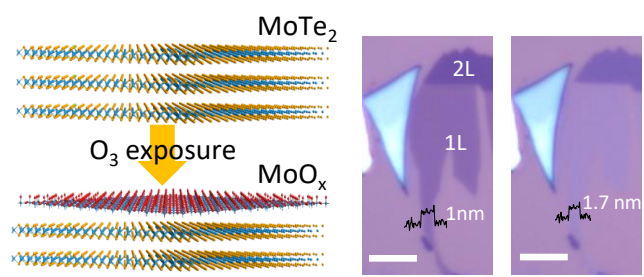

# Supporting Information

## Controlled layer-by-layer oxidation of MoTe$_2$ via O$_3$ exposure

Xiaoming Zheng[1,3,#], Yuehua Wei[2,#], Chuyun Deng[1], Han Huang[3], Yayun Yu[1], Guang Wang[1], Gang Peng[1], Zhihong Zhu[2], Yi Zhang[1], Tian Jiang[2], Shiqiao Qin[2], Renyan Zhang[2,*], Xueao Zhang[1,3,*]

[1]College of Arts and Science, National University of Defense Technology, Changsha, Hunan, 410073, China
[2]Academy of Advanced Studies in Interdisciplinary Research, National University of Defense Technology, Changsha, Hunan, 410073, China
[3]School of Physics and Electronics, Central South University, Changsha, Hunan, 410004, China

[#]X.Z and Y.W contribute equally to this work
[*]correspondence should be addressed to R.Z: ryancms@sina.cn and X.Z: xazhang@nudt.mtn

**This file includes:**

Experimental section
    *Sample preparation*
    *Characterization*
Supplementary Text
    *Transport measurement of oxidized monolayer MoTe$_2$*
    *Application of oxide film on MoTe$_2$:*
Figure S1 to S13

**Experimental section:**

*Sample preparation:*

In the experiment, atomically thin $MoTe_2$ flakes with different number of layers were mechanically exfoliated from the bulk crystals (supplied by HQ Graphene) onto silicon wafer with 300 nm oxide. The thickness of these flakes was identified by optical microscopy, which would be further analyzed by Raman spectroscopy and atomic force microscopy (AFM). All the samples were annealed in Ar with 10% $H_2$ at 250 °C for 2h. The layer-by-layer oxidation of $MoTe_2$ is achieved by introducing thin $MoTe_2$ flakes into a commercial ultraviolet (UV) – $O_3$ cleaning system (UV/Ozone Procleaner TM plus, Bioforce Nanoscience, USA) to expose to $O_3$ for various times at temperatures from room temperature to 200 °C. And the exposing time and temperature are dependent on the desired degree of oxidation. UV-light will split the oxygen molecule ($O_2$) into oxygen free radicals, then these oxygen free radicals hit another $O_2$ molecules to form $O_3$. The concentration of the generated $O_3$ gas is determined by the absorption of UV light at 254 nm[1,2] in a quartz flow gas cell through U-100 spectrophotometer (Hitachi, Japan). Figure S1 shows the typical UV absorption spectrum of $O_3$ gas in our experiment, from which the $O_3$ concentration is extracted roughly as 180 mg/m$^3$, which agrees quite well with the manual of UV-$O_3$ cleaning system. Detail of this method could be found in ref.[1,2].

*Characterization:*

$MoTe_2$ flakes were characterized by optical microscopy, AFM, Raman, XPS and back gate FET transport measurement before and after $O_3$ exposure. The optical images were taken by LV100D system (Nikon, Japan). AFM measurements were carried out in a Ntegra Prima system (NT-MDT, Russia), using a semi-contact mode.

Raman spectra were performed using Alpha 300R system (Witec Company, Germany) with a 100x objective lens in ambient. During Raman measurement, the laser power was kept below 0.1 mW to avoid any damages to $MoTe_2$ crystals. The exciting laser wavelengths are 532 and 633 nm with laser spot size of about 1 μm. The gratings of 1800 and 600 grooves/mm were adopted for Raman measurements.

To obtain XPS spectra, we used PHOIBOS 100 hemispherical energy analyzer (SPECS, Germany) with Al Kα micro-focused monochromatized source. For the XPS measurements, we used a bulk $MoTe_2$ crystal before and after $O_3$ exposure for 60 mins, with the analysis spot size of 600 μm. In this report, the binding energies were calibrated using C 1s peak at 284.6 eV as reference. Lorentzian-Ganssian peak fitting was performed for analyzing the data with standard Shirley baseline.

The $MoTe_2$ FET devices were defined by e-beam lithography (Raith e-LINE Plus, Germany) and then evaporated 5nm Ag and 50 nm Au as the contact using e-beam evaporation (Kurt J. Lesker PVD75, USA). Transport measurements were conduct using a Keithley 4200 semiconductor parameter analyzer integrated with a shielded probe station (CGO-4, Cindbest, China) in high vacuum at various temperature. Two-probe FET hole mobility and sheet density of holes of $MoTe_2$ flake can be evaluated from the linear plot of transfer curves on the p-side by the equation: $\mu_h = \frac{L}{W C_g V_d} \frac{dI_d}{dV_g}$ and $n_{2d} = \frac{C_g(V_g - V_{th})}{e}$, respectively, where L and W are the length and width of the channel, $C_g$ = 1.2 x 10$^{-8}$ F cm$^{-2}$ is the capacitance of the back gate SiO$_2$, $V_d$ is the source-drain voltage drop, $V_g$ is the gate voltage, $V_{th}$ is the threshold voltage, and e is the electron charge (Figure S10).

**Supplementary Text**

*Transport measurement of oxidized monolayer MoTe$_2$:*
The hole-doping is likely caused by local oxidation of underlying single layer MoTe$_2$ and/or hole injection from top high working function oxide film (MoO$_x$, with x<3)[3-5]. However, we rule out the possibility of local oxidation by transport measurement of oxidized monolayer MoTe$_2$. We measured the $I_d$ - $V_d$ and $I_d$ - $V_g$ characteristics of monolayer MoTe$_2$ before and after O$_3$ intermittent exposure from 0 to 60 s (Figure S11). It is clear that the conductivity is significantly reduced with O$_3$ exposure and the MoO$_x$ film is insulating. The insulating properties is also found in heavily oxidized thicker MoTe$_2$ samples, for example the fully oxidized trilayer MoTe$_2$ in Figure S13. This excludes the possibility that the high p-type doping come from the local oxidation of single layer MoTe$_2$ and that the MoO$_x$ itself is hole dominate conductor. Thus, we attribute the nearly metallic conduction to the heavily p-doped underneath MoTe$_2$. Actually, the p-type doping effect has already been demonstrated by Raman and XPS measurements. Since the oxide is nucleated from the edge as well as surface defects and grows laterally, thus MoTe$_2$ should be oxidized homogeneously even beneath metal electrodes independent of the lateral size, as already been seen in WSe$_2$[6]. This thin oxide layer of MoO$_x$ between the contact and MoTe$_2$ allows to unpin their Fermi level at the metal contact, which serves as an Ohmic or low-Schottky barrier contact. Thus, the tunneling of charge carriers through the contact barriers become more effective, inducing orders of increase of the two-probe FET mobility.

*Application of oxide film on MoTe$_2$:*
It is instructive to show two possible applications of surface oxide in MoTe$_2$. Firstly, the surface MoO$_x$ layer could act as a passivation layer for MoTe$_2$. As is known to all, MoTe$_2$ is acutely sensitive to air exposure[7]. Figure S7a shows that the Raman signal of pristine monolayer disappear in about 100 h. However, the influence of air exposure for MoO$_x$ covered MoTe$_2$ is hardly detectable even after 400 h (Figure S7b). Secondly, it could be used as a good seed layer for ALD of high dielectrics, such as Al$_2$O$_3$. Due to the absence of dangling bonds, MoTe$_2$ turns out to be difficult to grow thin Al$_2$O$_3$ film as in Figure S8a. Although previous researches show that atomically thin MoO$_x$ film is not a good dielectrics[6], however, after oxidation, it is become much easier to deposit homogenous thin Al$_2$O$_3$ film (less than 30 nm, as in Figure S8b). In addition, surface MoO$_x$ film could also serve as a protective layer to prevent the underneath MoTe$_2$ layer from physical bombardment during deposition process.

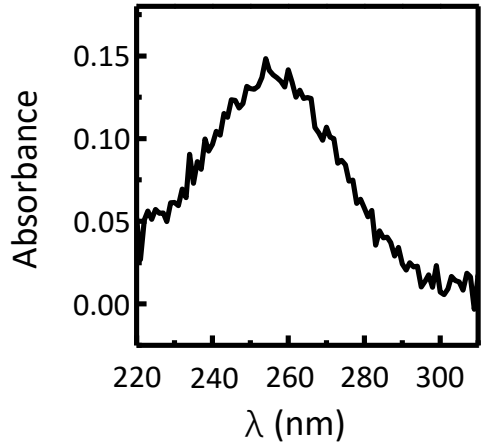

**Figure S1|** UV absorption spectrum of $O_3$. The extract $O_3$ density is 180 mg/m$^3$.

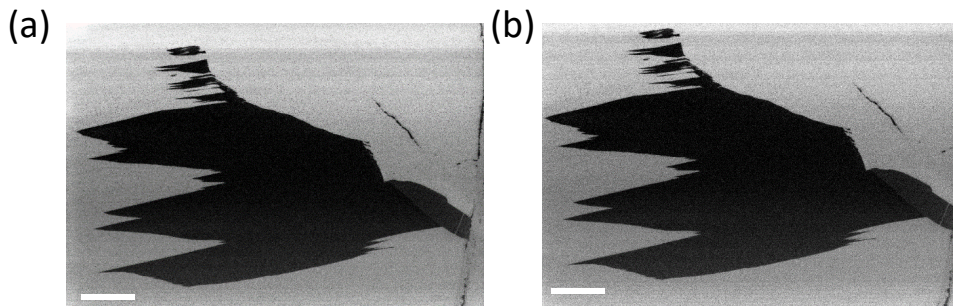

**Figure S2|** SEM image of monolayer MoTe$_2$ before (a) and after (b) $O_3$ exposure for 2 mins. The scale bar is 1 μm.

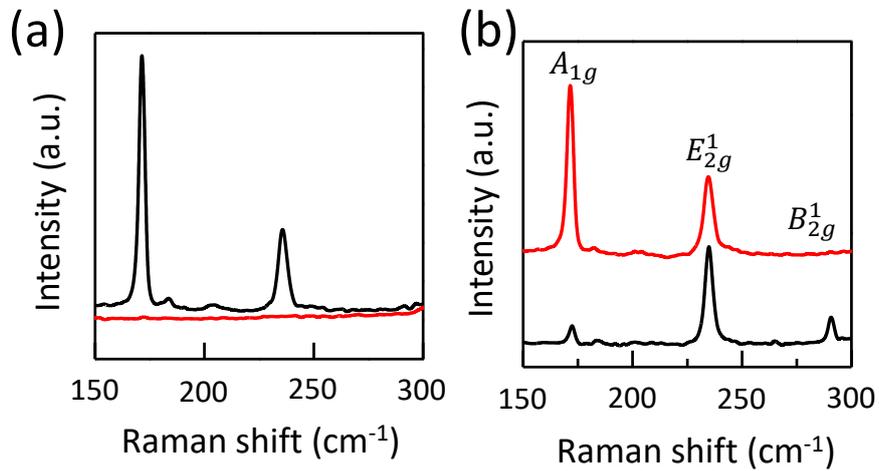

**Figure S3|** Raman spectra of monolayer (a) and bilayer (b) MoTe$_2$ before (black curves) and after (red curves) $O_3$ exposure for 2 mins with an exciting laser wavelength of 633 nm.

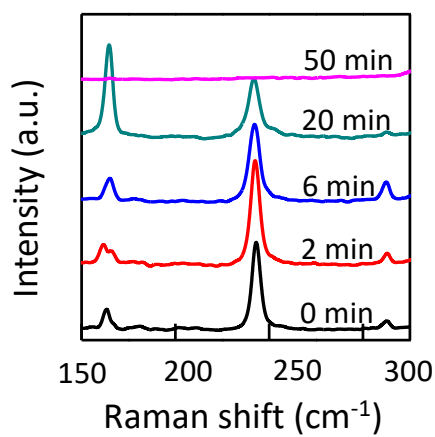

**Figure S4|** Raman spectra (laser wavelength 633 nm) of tetralayer MoTe$_2$ upon intermittent O$_3$ exposure from 0 to 50 mins.

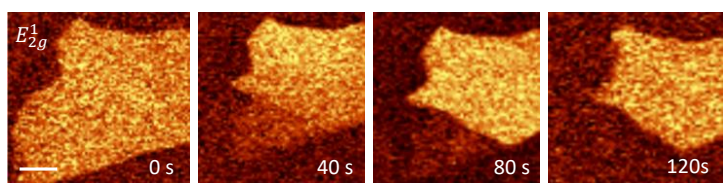

**Figure S5|** Raman mapping of $E^1_{2g}$ mode of bilayer and monolayer MoTe$_2$ upon intermittent O$_3$ exposure from 0 to 120 s. The scale bar is 5 μm.

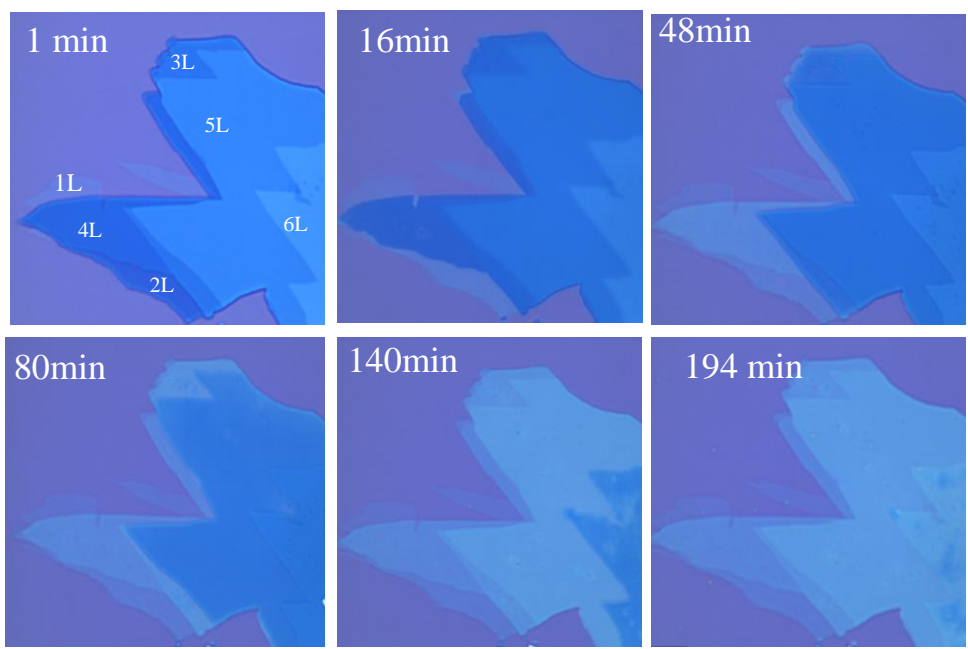

**Figure S6|** Layer-by-layer oxidation of few layer MoTe$_2$. Optical images of six-layer MoTe$_2$ upon intermittent O$_3$ exposure from 0 to 194 mins.

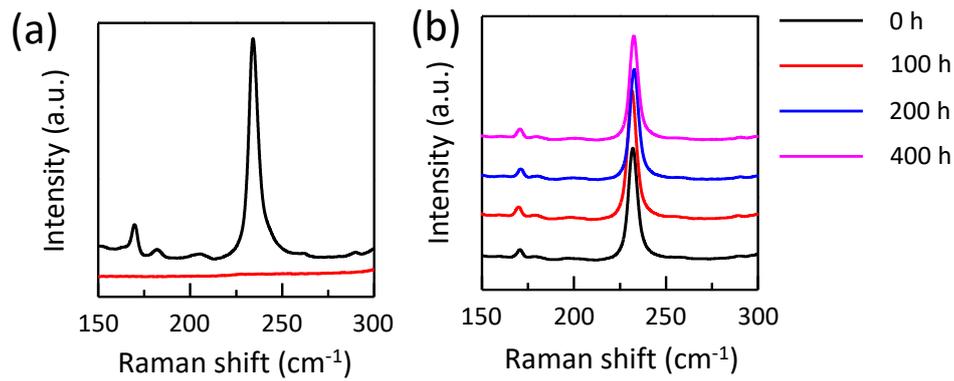

**Figure S7|** Passivation of MoTe$_2$ flakes after O$_3$ exposure. (a, b) Raman spectra evolution versus air exposure time of pristine monolayer and oxidized bilayer MoTe$_2$ from 0 to 400 h, respectively.

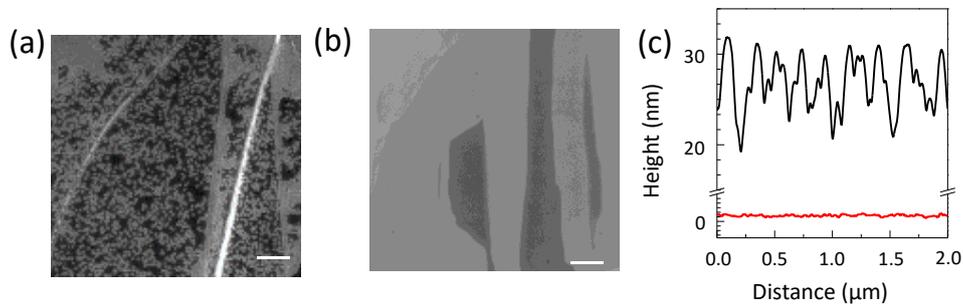

**Figure S8|** Seed layer for ALD high dielectrics Al$_2$O$_3$. (a, b) SEM image of 30 nm Al$_2$O$_3$ on pristine and oxidized MoTe$_2$ flakes, respectively. The scale bar is 1 μm. (c) Corresponding height profile for Al$_2$O$_3$ on pristine (black curve) and O$_3$ exposure (red curve) MoTe$_2$ flakes.

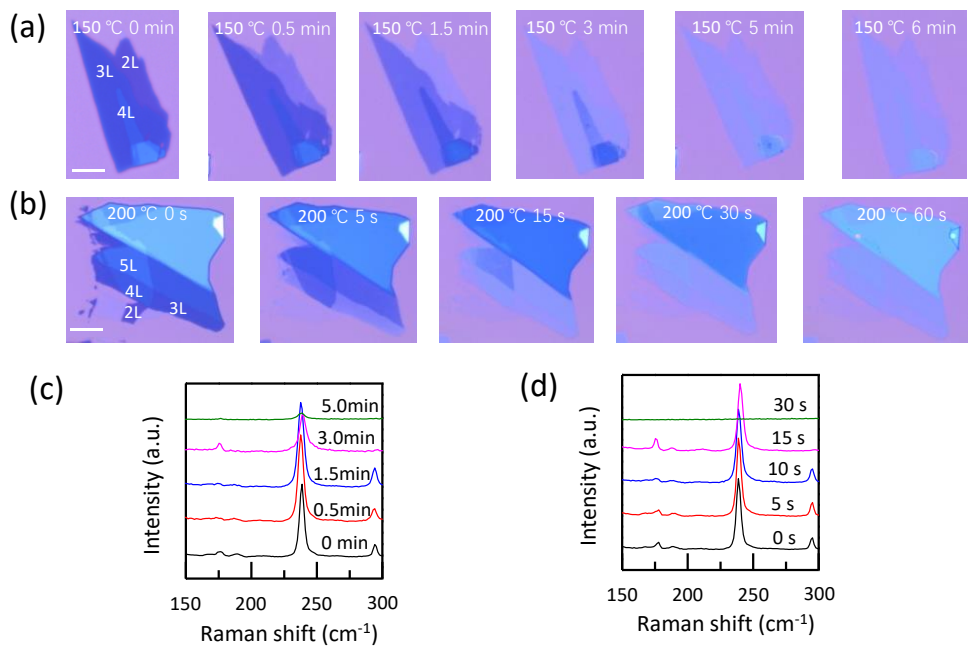

**Figure S9|** (a, b) Optical image of MoTe$_2$ flakes upon intermittent O$_3$ exposure from 0 to 6 mins at 150 ℃ and 0 to 60 s at 200 ℃, respectively. The scale bars are 5 μm. (c) Raman spectra

of tetralayer MoTe$_2$ flakes upon intermittent O$_3$ exposure from 0 to 5 mins at 150 ℃. (d) Raman spectra of five-layer MoTe$_2$ flakes upon intermittent O$_3$ exposure from 0 to 30 s at 200 ℃.

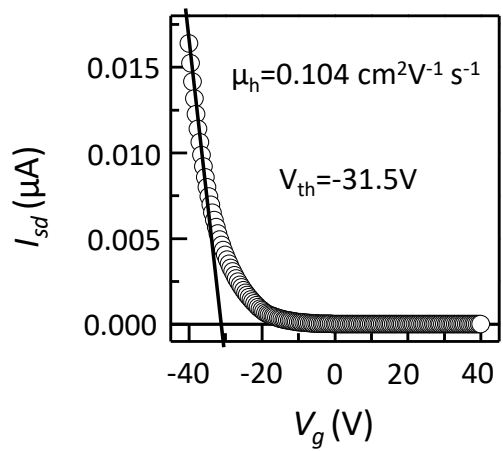

**Figure S10|** Determination of the two-probe FET hole mobility and carrier density. Pristine bilayer MoTe$_2$ transistor is taken as an example. $\frac{dI_d}{dV_g}$ is given by the linear region of the transfer characteristic $I_d$ - $V_g$ plot. $V_{th}$ is the threshold voltage

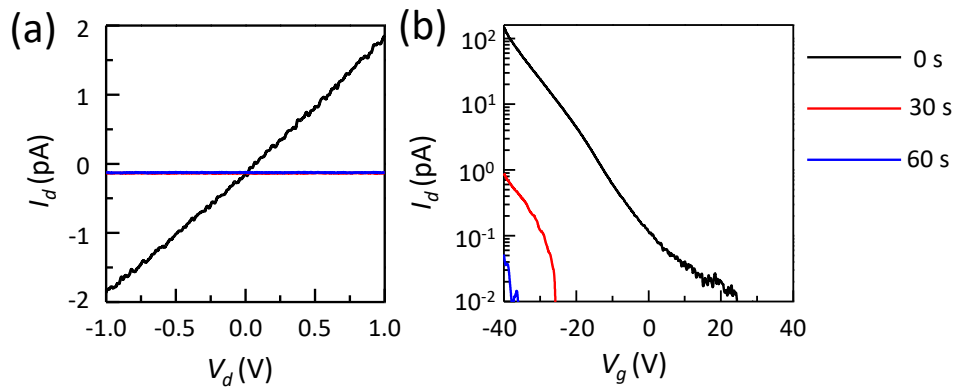

**Figure S11|** (a, b) $I_d$ - $V_d$ and $I_d$ – $V_g$ characteristics of monolayer MoTe$_2$ transistor upon intermittent O$_3$ exposure from 0 to 60 s, respectively.

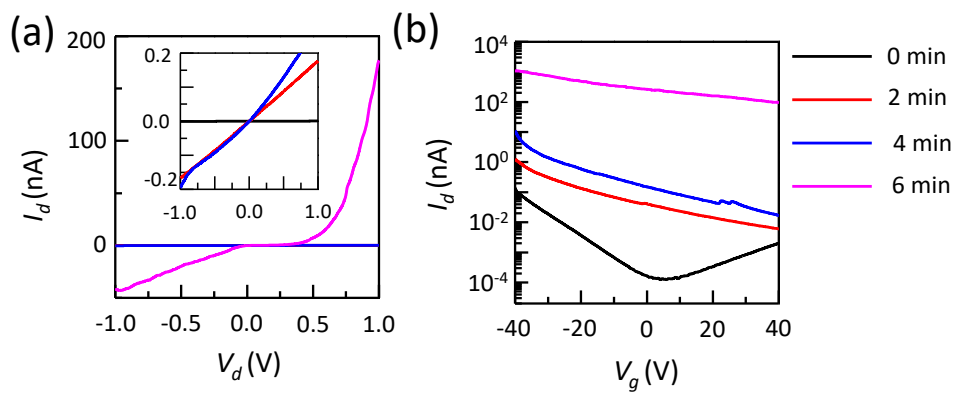

**Figure S12|** (a, b) $I_d$ - $V_d$ and $I_d$ – $V_g$ characteristics of trilayer MoTe$_2$ transistor upon intermittent O$_3$ exposure from 0 to 6 mins, respectively.

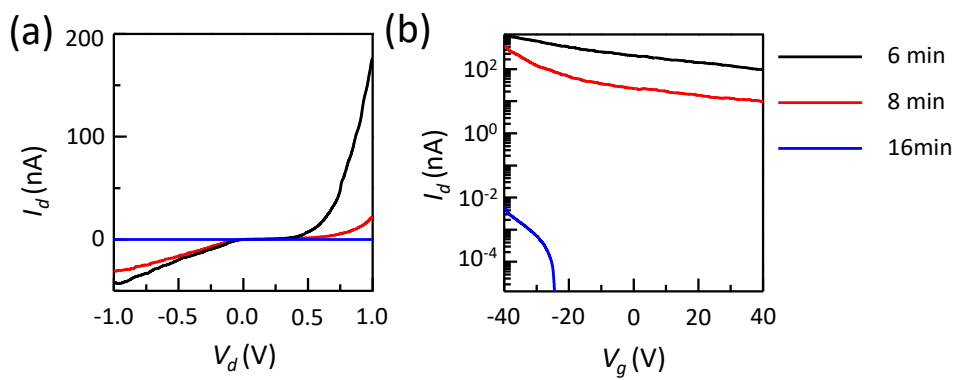

**Figure S13|** (a, b) $I_d$ - $V_d$ and $I_d$ – $V_g$ characteristics of trilayer MoTe$_2$ transistor upon intermittent O$_3$ exposure from 6 to 16 mins, respectively.

**References**

(S1) McElroy, F., Mikel, D., Nees, M. , *McElroy F, Mikel D, Nees M. Determination of ozone by ultraviolet Analysis, A new method for Volume II, Ambient Air Specific Methods[J]. Quality Assurance Handbook for Air Pollution Measurement Systems. URL: http://mattson. creighton. edu/Ozone/OzoneEPAMethod. pdf (Accessed 9 October 2011),* 1997.

(S2) Malicet, J.; Daumont, D.; Charbonnier, J.; Parisse, C.; Chakir, A.; Brion, J., Ozone UV Spectroscopy. II. Absorption Cross-Sections and Temperature Dependence. *J. Atmos. Chem.* 1995, 21, 263-273.

(S3) Meyer, J.; Hamwi, S.; Kröger, M.; Kowalsky, W.; Riedl, T.; Kahn, A., Transition Metal Oxides for Organic Electronics: Energetics, Device Physics and Applications. *Adv. Mater.* 2012, 24, 5408-5427.

(S4) Greiner, M. T.; Lu, Z. H., Thin-Film Metal Oxides in Organic Semiconductor Devices: their Electronic Structures, Work Functions and Interfaces. *Npg Asia Mater.* 2013, 5, 547-556.

(S5) Michaelson, H. B., The Work Function of the Elements and Its Periodicity. *J. Appl. Phys.* 1977, 48, 4729-4733.

(S6) Yamamoto, M.; Nakaharai, S.; Ueno, K.; Tsukagoshi, K., Self-Limiting Oxides on WSe$_2$ as Controlled Surface Acceptors and Low-Resistance Hole Contacts. *Nano Lett.* 2016, 16(4): 2720-2727.

(S7) Chen, B.; Sahin, H.; Suslu, A.; Ding, L.; Bertoni, I.; Peeters, M.; Tongay, S., Environmental Changes in MoTe$_2$ Excitonic Dynamics by Defects-Activated Molecular Interaction. *ACS Nano* 2015, 9, 5326-5332.